\documentclass[12pt]{article}
\usepackage{rotating}
\usepackage{epsfig,amssymb}
\usepackage{ulem}
\input{epsf}

\def\laq{\raise 0.4 ex \hbox{$<$}\kern -0.8 em\lower 0.62 ex\hbox{$\sim$}}
\def\gaq{\raise 0.4 ex \hbox{$>$}\kern -0.7 em\lower 0.62 ex\hbox{$\sim$}}

\def\beq{\begin{equation}}
\def\eeq{\end{equation}}
\def\beqa{\begin{eqnarray}}
\def\eeqa{\end{eqnarray}}

\def\ph{\phi}

 \def\frac#1#2{{\textstyle{{#1}\over {#2}}}}
 \def\lsim{\mathrel{\rlap{\lower4pt\hbox{\hskip1pt$\sim$}}
    \raise1pt\hbox{$<$}}} \def\gsim{\mathrel{\rlap{\lower4pt\hbox{\hskip1pt$\sim$}}
    \raise1pt\hbox{$>$}}}
\def\sqr#1#2{{\vcenter{\vbox{\hrule height.#2pt
         \hbox{\vrule width.#2pt height#1pt \kern#1pt
         \vrule width.#2pt}
         \hrule height.#2pt}}}}

\def\gappeq{\mathrel{\rlap {\raise.5ex\hbox{$>$}} {\lower.5ex\hbox{$\sim$}}}}

\def\lappeq{\mathrel{\rlap{\raise.5ex\hbox{$<$}}
{\lower.5ex\hbox{$\sim$}}}}

\usepackage{color}

\begin{document}
\pagestyle{plain}

\begin{flushright}
May 3, 2022
\end{flushright}
\vspace{15mm}

\begin{center}

{\Large\bf Multi-field cold and warm inflation and the de Sitter swampland conjectures }

\vspace*{1.0cm}

Orfeu Bertolami$^{1,2}$ and Paulo M. S\'a$^{3,4}$ \\
\vspace*{0.5cm}
{$^{1}$ Departamento de F\'{\i}sica e Astronomia, Faculdade de Ci\^encias,
Universidade do Porto, \\
Rua do Campo Alegre s/n, 4169-007 Porto, Portugal}\\

{$^{2}$ Centro de F\'{\i}sica das Universidades do  Minho e do Porto,
Rua do Campo Alegre s/n, 4169-007 Porto, Portugal}\\

{$^{3}$ Departamento de F\'{\i}sica, Faculdade de Ci\^encias e Tecnologia,
Universidade do Algarve, \\
Campus de Gambelas, 8005-139 Faro, Portugal}\\

{$^{4}$ Instituto de Astrof\'{\i}sica e Ci\^encias do Espa\c co,
Faculdade de Ci\^encias, Universidade de Lisboa,
Campo Grande, 1749-016 Lisboa, Portugal}

\vspace*{2.0cm}
\end{center}

\begin{abstract}
\noindent
We discuss under which conditions multi-field cold and warm inflationary
models with canonical kinetic energy terms are compatible with the swampland 
conjectures about the emergence of de Sitter solutions in string theory.
We find that under quite general conditions the slow-roll conditions for
multi-field cold inflation are at odds with the swampland conjectures for
an arbitrary number of scalar fields driving inflation. However, slow-roll
conditions can be reconciled with the swampland conjectures in the strong
dissipative regime of warm inflation. 
\end{abstract}

\vfill
\noindent\underline{\hskip 140pt}\\[4pt]
\noindent
{E-mail addresses: orfeu.bertolami@fc.up.pt; pmsa@ualg.pt}

\newpage
\section{Introduction}
\label{sec:intro}

\phantom{.}\indent Despite its impressive capability to comprise the most
desirable features of a fundamental theory of quantum gravity, 
string theory has failed, among other shortcomings, to allow for a natural
scenario for inflation.
This situation arises as the fundamental scalar field of the theory, 
the dilaton, does not acquire any potential in any order of perturbation
theory and non-perturbative contributions tend to lead to very shallow and
phenomenologically untenable models.
In fact, de Sitter solutions seem to be hard to obtain in string theory,
even though possible routes have been proposed  \cite{KKLT}. 
This is sharply contrasting with the situation of $N=1$ supergravity in the
context of which successful inflationary models can be rather easily achieved
\cite{Adams}.
This is somewhat surprising as, in fact, it is thanks to inflation that,
for instance, some specific issues of superstring models with intermediate
scale can be circumvented \cite{OBRoss87}.
Of course, beyond single-field inflation, proposals for two-field inflation 
have been put forward (see e.g.\ \cite{Linde94,Bento91}), but to our
knowledge these have not been systematically explored in the context of
string theory.

The recent set of conjectures that de Sitter solutions in string theory
cannot be found in its landscape, but lye actually in a ``swampland",
i.e.\ the set of consistent-looking theories that do not admit a suitable
ultraviolet completion in string theory, has led to an intense discussion
whether the resulting solutions are consistent with the conditions required
by inflation.
These set of conjectures comprise quite general criteria: charge symmetries are
supposed to be local gauge symmetries so that at least one particle must have
a mass in Planck units less than the gauge coupling strength so to ensure that
gravity is weak; the sign of some higher-order terms in the effective action
is constrained to warrant the absence of superluminal propagation; there are
a finite number of massless particles (see Ref.~\cite{Palti} for a review).
More specifically, in what concerns inflation, these conjectures 
can be expressed in terms of constraints on 
scalar fields, generically denoted by $\phi$ in the field space \cite{OV,OOSV}:
\beq
{\Delta \phi \over M_{\rm P}} < c_1, 
\label{eq:SC1}
\eeq
\beq
M_{\rm P}{|\partial_\phi V| \over V} > c_2, 
\label{eq:SC2}
\eeq
where $\Delta\phi$ is the range of variation of the field, 
$M_{\rm P}\equiv M_{\rm Pl}/\sqrt{8\pi}$ is the Reduced Planck's mass,
$V(\phi)$ is the scalar field potential,
$\partial_\phi V \equiv \partial V/ \partial\phi$,
and $c_1$ and $c_2$ are $\mathcal{O}(1)$ constants.
It is argued that one should also  consider a more refined condition 
\cite{OPSV,garg-krishnan,Andriot}
\beq
M_{\rm P}^2{\partial_{\phi\phi}^{2} V \over V} < -c_3, 
\label{eq:SC3}
\eeq
where $\partial_{\phi\phi}^{2} V \equiv \partial^2 V/\partial\phi^2$
and, likewise $c_1$ and $c_2$, the constant $c_3$ is of order one.

Conditions given by Eqs.~(2) and (3) are somewhat at odds with
the onset conditions of single-field inflation which require that the
parameters for the inflaton field \cite{PDG2020},
\beq
\epsilon_{\phi}= {M_{\rm P}^2 \over 2} \left(
{\partial_\phi V  \over V} \right)^2
\label{eq:epsilon}
\eeq
and
\beq
\eta_{\phi}=M_{\rm P}^2  {\partial_{\phi\phi}^{2}
 V\over V},
\label{eq:eta}
\eeq
satisfy the slow-roll requirements, $\epsilon_{\phi} \ll 1$ and
$|\eta_{\phi}| \ll 1$ at the onset of inflation, so that at the end of
inflation $\epsilon_{\phi} \sim |\eta_{\phi}| \sim1$.
In order to successfully solve the initial conditions problems of standard
cosmology, the number of 
e-foldings of inflation must satisfy the condition $N_e >65$,
where
\beq
 N_e \equiv \ln\left({a_e \over a_i}\right) 
 \simeq - {1 \over M_{\rm P}^2} \int_{\phi_i}^{\phi_e} 
 	{V \over \partial_\phi V} d\phi
 = - {1 \over \sqrt{2} M_{\rm P}}\int_{\phi_i}^{\phi_e}
 	{1 \over \sqrt{\epsilon_{\phi}}}d\phi,
\label{eq:N_e}
\eeq
with $\phi_i$ and $\phi_e$ corresponding to the initial and final values of
the inflaton field. Since during inflation the slow-roll parameters are
approximately constant,
$d\epsilon_{\phi}/dN \simeq \mathcal{O}(\epsilon_{\phi}^2)$, hence
\beq
N_e  \simeq  {1\over\sqrt{2\epsilon_{\phi}}}   {\Delta \phi \over M_{\rm P}},
\eeq
and thus for $\epsilon_{\phi} \lesssim 10^{-4}$
one obtains $\Delta \phi \sim M_{\rm P}$,
which is consistent with the constraint given by Eq.~(\ref{eq:SC1}).

Further contact with observations can be established through the amplitude
of the inflaton quantum fluctuations and its imprint of the Cosmic Microwave
Background Radiation (CMB) through curvature
and density fluctuations,
\beq
\Delta_R^2 = A_s  \left({k \over k_*}\right)^{n_s-1},
\eeq
where $k_*$ is a chosen scale measured on the CMB, the scalar spectral index is
\beq
n_s \simeq 1 - 6 \epsilon_{\phi} + 2 \eta_{\phi},
\eeq
and the amplitude of scalar fluctuations is given by 
\beq
A_s  \simeq {1 \over 24 \pi^2} \left(
{V(\phi_H) \over M_{\rm P}^4} \right )
{1 \over \epsilon_{\phi} (k_H)},
\eeq
with $\phi_H$ and $\epsilon_{\phi}(k_H)$ corresponding
to a scale when the fluctuating modes cross the horizon,
that is between 50 and 60 e-foldings before the end of inflation.

Furthermore, from the spectrum of tensor modes generated by the
inflaton quantum fluctuations,
\beq
\Delta_T^2 = A_t  \left({k \over k_*}\right)^{n_t},
\eeq
where
\beq
A_t \simeq {2 \over 3 \pi^2} \left({V(\phi_H) \over M_{\rm P}^4}\right )
\eeq
and
\beq
n_t \simeq -2 \epsilon_{\phi},
\eeq
it is found that the tensor-to-scalar ratio is given by
\beq
r= {\Delta_T^2 \over \Delta_R^2} \simeq 16 \epsilon_{\phi}.
\eeq

Planck 2018 temperature and polarisation data indicate that for
$0.008 \, h^{-1} {\rm Mpc}^{-1} \leq k \leq 0.1 \, h^{-1} {\rm Mpc}^{-1}$, 
the expansion rate $H = (67.4 \pm 0.5) \,{\rm km\, s}^{-1} {\rm Mpc}^{-1}$,
and absence of running, the scalar spectral index is 
\beq
n_s=0.9649 \pm 0.0042,
\eeq
while BICEP2/Keck Array, Planck and other data place an upper bound on the
tensor-to-scalar ratio at $k = 0.002 \, {\rm Mpc}^{-1}$, namely,
\beq
r < 0.06.
\eeq
From which follows, dropping higher-order
slow-roll parameters, that
\beq
\epsilon_{\phi} <0.0044
\label{eq:ePlanck}
\eeq
and
\beq
\eta_{\phi} = - 0.015 \pm 0.006,
\label{eq:etaPlanck}
\eeq
which do not match $c_2$ and $c_3$. 

In what follows we shall consider the situation in the context of
multi-field cold inflationary models with canonical kinetic energy terms,
initially for two and then for an arbitrary number of fields
(section~\ref{sec:cold}).
We shall then address in section~\ref{sec:warm} the problem in the context
of warm inflationary models which exhibit dissipation.
We shall see that the swampland conjectures and the slow-roll conditions
can be reconciled in
the strong dissipative regime of warm inflation
for one, two, and multi-field models.
Conclusions will be drawn in section~\ref{sec:concl}.

\section{Beyond Single Field inflation}

\phantom{.}\indent As seen above, the de Sitter swampland conjectures and
the slow-roll conditions cannot be matched for single-field cold
inflation.
As discussed, the purpose of the swampland conjectures is to ensure that a
suitable effective field theory in a de Sitter background arises from
string theory.
It is further assumed that these conjectures also allow for the emergence of
a classical theory of gravity, which then drives inflation.
The connection between curvature and matter-energy as established by General
Relativity (GR) is well supported experimentally and observationally
(see e.g.\ \cite{Will,BP2014} for discussions).
Indeed, theoretical and experimental evidence suggest that GR reflects a
provisional stage, although highly relevant, towards the ultimate description
of gravity. Well-known theoretical difficulties concern the existence of
spacetime singularities and the cosmological constant problem, both related 
to the key issue of making quantum mechanics and GR compatible with each other, 
string theory being, of course, quite relevant in this respect.  

It is quite logical that de Sitter swampland conjectures have attracted
great interest of the cosmological community given that they impose
conditions on the field spacing and on the first two derivatives of the
potential of the background effective field theory (see, for instance, 
Ref.~\cite{Andriot} for a partial list), which can be accessed
in single-field inflation with features of the CMB
as discussed above.

Clearly, the conjectures~(\ref{eq:SC2}) and (\ref{eq:SC3}) are at
odds with constraints arising from
the CMB data, Eqs.~(\ref{eq:ePlanck}) and (\ref{eq:etaPlanck}).
This conflict has been pointed out, for instance, in Ref.~\cite{Kinney},
even though it has been argued that definite conclusions about an actual tension 
depend on the knowledge about the origin of the adiabatic
curvature perturbations, within the slow-roll 
single-field models of inflation \cite{Riotto}.
In fact, the incompatibility of the swampland conjectures with the
observations has been an object of critique from the authors of
Ref.~\cite{Linde2019}.  
In any case, the swampland conjectures have given origin to many ideas
and sparked interesting proposals
\cite{Geng2020,Wang2020,Mavromatos2020,Grana2021,Cicoli2021}.

Even though the conflict is still depending on the nature of the
perturbations, a natural way to avoid this tension is to consider
multi-field inflationary models.
Most of these models  are known to show no contradiction with the CMB
features \cite{Wands}.
In fact, multi-field models open interesting perspectives, for instance,
for unification with dark matter and dark energy \cite{Sa2009,Sa2020}.
Two-field inflationary models were first considered in the context of $N=1$
supergravity \cite{Ovrut} and their dynamics was scrutinized in
Refs.~\cite{OBRoss86,OB88} for a broader class of models.
In a broad context and in string theory, 
two-field inflationary with different mass scales and an
interaction term were considered in Refs.~\cite{Linde94,Bento91,Bento92}.
In the context of the swampland conjectures, two-field inflationary models 
were discussed in 
Refs.~\cite{Achucarro,Gashti}, where in Ref.~\cite{Achucarro}
non-canonical kinetic energy terms have been considered.
However, in what follows, we shall present quite general arguments
which seem to rule out a putative conciliation using multi-field cold 
inflation with canonical kinetic energy terms (see section~\ref{sec:cold}).
We shall consider first a two-scalar cold inflationary model and then generalize
the argument for multi-field inflationary models. 
In section~\ref{sec:warm} we turn to the case of warm inflation, for which,
in the strong dissipative regime, slow-roll conditions can be reconciled with 
the swampland conjectures.

\subsection{Cold inflation\label{sec:cold}}

\phantom{.}\indent Quite generically, inflationary models driven by
two scalar fields $\phi$ and $\chi$
can be described by the Lagrangian density
\beq
{\cal L } = - {1\over2} \partial_{\mu} \phi \partial^{\mu} \phi
	- {1 \over 2} \partial_{\mu} \chi \partial^{\mu} \chi
	- U(\phi, \chi),
\label{B2}
\eeq
where the potential $U(\phi, \chi)$ contains the potentials of the fields
$\phi$ and $\chi$ and an interaction term between these fields. 
This potential is assumed to be designed to yield a healthy period of inflation.
Let us also assume that it satisfies
the conjecture~(\ref{eq:SC1}),
even though it might be trickier to achieve this condition for many fields. 
But, as we shall see, if conjecture~(\ref{eq:SC1}) can, at least in
principle, be satisfied, the conjecture~(\ref{eq:SC2}) cannot be met
in the context of quasi-exponential cold
inflation. 

Coupling these fields to gravity in an homogeneous, isotropic and flat
cosmological spacetime background and assuming that the scalar fields
can decay to radiation leads to the field equations
\beq
\ddot{\phi}+3H\dot{\phi} + \partial_{\phi} U =
	 -\Gamma_\phi \dot{\phi},
\label{B3}
\eeq
\beq
\ddot{\chi}+3H\dot{\chi} + \partial_{\chi} U =
	-\Gamma_{\chi} \dot{\chi}, 
\label{B4}
\eeq
\beq
\dot{\rho}_{\rm R} + 4H\rho_{\rm R} = \Gamma_{\phi}  {\dot \phi}^2
+ \Gamma_{\chi} {\dot \chi}^2,
\label{Br}
\eeq
\beq
 H^2= {1 \over 3M_{\rm P}^2}\left({1 \over 2} {\dot{\phi}^2}
 + {1 \over 2} \dot{\chi}^2+U+\rho_{\rm R} \right), 
 \label{B1}
 \eeq
 \beq
 \dot{H}=-{1 \over 2M_{\rm P}^2} \left( \dot{\phi}^2+\dot{\chi}^2 +
 {4\over 3}\rho_{\rm R} \right),
 \label{B2}
\eeq
where, in general terms, $\Gamma_{\phi} = \Gamma_{\phi}(\phi)$ and
$\Gamma_{\chi}=\Gamma_{\chi}(\chi)$
denote the decay widths of these fields to radiation, $\rho_{\rm R}$ is the
energy density of the radiation fluid, $H=\dot{a}/a$ is the expansion rate,
and $a$ the scale factor of the Friedman--Lema\^{\i}tre--Robertson--Walker
metric.
In the above equations, an overdot denotes a derivative with respect to the
cosmic time $t$.

In this subsection we shall consider that $\Gamma_{\phi}=\Gamma_{\chi}= 0$
and neglect $\rho_{\rm R}$, meaning that inflation will supercool the Universe
and a reheating process must be considered separately.
Non-vanishing dissipative terms and a non-negligible energy density of radiation
will be considered in section~\ref{sec:warm}.

Let us assume that inflation is almost exponential, namely,
\begin{equation}
{|\dot{H}| \over H^2} \ll1. 
 \label{B5}
\end{equation}

From this expression, using Eqs.~(\ref{B1}) and (\ref{B2}), follows that
$\dot{\phi}^2+\dot{\chi}^2\ll U$, which implies that Eq.~(\ref{B1})
can be written as
\begin{equation}
U\simeq 3M_{\rm P}^2 H^2.
 \label{B6}
\end{equation}
In the above expression and in what follows, the symbol $\simeq$ means
``equal within the slow-roll approximation".

Taking the time derivative of Eq.~(\ref{B6}) and using Eq.~(\ref{B2}) we obtain
\begin{equation}
\dot{\phi} \partial_\phi U + \dot{\chi} \partial_\chi U 
\simeq - 3H (\dot{\phi}^2+\dot{\chi}^2), 
\label{B7}
\end{equation}
which can be written as
\begin{equation}
\partial_\phi U + 3H\dot{\phi} \left(1+
{\dot{\chi}^2 \over \dot{\phi}^2} \right)
 \left( 1+ {\dot{\chi}\partial_\chi U \over \dot{\phi}\partial_\phi U}
 \right)^{-1} \simeq 0. 
\label{B8}
\end{equation}
Imposing the condition
\begin{equation}
\left(1+{\dot{\chi}^2 \over \dot{\phi}^2}\right)
\left( 1+ {\dot{\chi}\partial_\chi U \over \dot{\phi}\partial_\phi U}
\right)^{-1} \simeq 1, 
\label{B9}
\end{equation}
we obtain
\begin{equation}
\partial_\phi U+3H\dot{\phi}\simeq0, 
\label{B10}
\end{equation}
which is equivalent to neglecting $\ddot{\phi}$ in Eq.~(\ref{B3}).

Now, using Eq.~(\ref{B10}), the condition given by Eq.~(\ref{B9})
can be simplified to yield
\begin{equation}
\partial_\chi U +3H\dot{\chi}\simeq 0, 
\label{B11}
\end{equation}
which is equivalent to neglecting $\ddot{\chi}$ in Eq.~(\ref{B4}).

Following Ref.~\cite{lyth-liddle},
the slow-roll parameter $\epsilon$ is defined as
\begin{equation}
\epsilon = {1 \over 2} M_{\rm P}^2 {|\nabla U|^2 \over U^2}, 
\label{B12}
\end{equation}
which is a straightforward generalization of the case with a single 
scalar field [see Eq.~(\ref{eq:epsilon})].
Using Eqs.~(\ref{B2})--(\ref{B6}), (\ref{B10}), and (\ref{B11}), we obtain
\beq
\epsilon \simeq {|\dot{H}| \over H^2} \ll 1. \\
 \label{B13}
\eeq

The de Sitter swampland conjecture
[see Eq.~(\ref{eq:SC2})]
\begin{equation}
 M_{\rm P} {|\partial_\phi U| \over U}>c_2
 \label{B14a}
\end{equation}
can be written as
\begin{equation}
c_2^2 \lesssim 2 {|\dot{H}| \over H^2} f, \\
 \label{B14}
\end{equation}	
where
\begin{equation}
f(\dot{\phi},\dot{\chi}) \equiv {\dot{\phi}^2 \over \dot{\phi}^2+\dot{\chi}^2}.
 \label{B15}
\end{equation}
Since the above function satisfies the condition
$0<f(\dot{\phi},\dot{\chi})\leq1$, hence we conclude that $c_2^2\ll1$.

The main issue is that, if we assume that inflation is quasi-exponential,
i.e., $|\dot{H}|/H^2\ll1$, then both $\epsilon$ and $c_2$ are much 
smaller than one.

These results can be straightforwardly generalized to 
the case of an arbitrary number of scalar fields. 
Indeed, let us consider the scalar fields $\phi$ and
$\chi_i$ ($i = 1, \dots, N$) with potential $U=U(\phi,\chi_1,\dots,\chi_N)$.

Once again, assuming quasi-exponential inflation and using the evolution 
equations --- generalized to the case of the above $N+1$ scalar fields 
--- we obtain
\begin{equation}
\sum_{i=1}^N \left( \dot{\chi_i} \partial_{\chi_i} U + 3 H \dot{\chi}_i^2
 \right) \simeq 0.
  \label{CI-f1}
\end{equation}
Since this equation \textit{does not} imply 
\begin{equation}
\partial_{\chi_i} U +3H\dot{\chi_i}\simeq 0, \quad (i=1,\dots,N),
\label{C11}
\end{equation}
these latter expressions must be assumed (as in Ref.~\cite{lyth-liddle}),
instead of being derived from Eq.~(\ref{B6}) as in the two-scalar field case.
The parameter $\epsilon$ satisfies the condition given by Eq.~(\ref{B13}), 
i.e., it is much smaller than unity during the inflationary period, and
$c_2^2$ is given by Eq.~(\ref{B14}), but now
\begin{equation}
f(\dot{\phi},	 \dot{\chi_1},\dots,\dot{\chi_N}  )
\equiv {\dot{\phi}^2 \over \dot{\phi}^2
+\sum_{i=1}^N\dot{\chi_i}^2}.
\label{C15}
\end{equation}
Since the above function satisfies the condition
$0<f(\dot{\phi},\dot{\chi_1},\dots,\dot{\chi_N}  )\leq1$,
it then follows that $c_2^2\ll1$, 
as in the two-scalar-field case.

Hence, it has been shown that, under the quite general conditions of cold
exponential inflation, the swampland conjectures cannot be met for any number
of scalar fields.

\subsection{Warm Inflation\label{sec:warm}} 

\phantom{.}\indent In the warm inflation scenario, dissipation plays a
crucial role in slowing down the inflaton, $\phi$, as it rolls down the
potential $V(\phi)$ \cite{Berera}.  Within a strong dissipative regime, it is known
that the swampland conjectures can be made compatible with 
single-field warm inflation \cite{mkr,bkr}.
In what follows, we show that this result extends to multi-field 
warm inflationary models. 

To characterize the slow-roll regime for a single field scenario,
in addition to the parameters $\epsilon_\phi$ and $\eta_\phi$,
defined by Eqs.~(\ref{eq:epsilon}) and (\ref{eq:eta}),
an additional parameter $\beta_\phi$ should be introduced,
\begin{equation}
 \beta_\phi = M_{\rm P}^2 \,
 {\partial_\phi\Gamma_{\phi}\over\Gamma_{\phi}} \,
 {\partial_\phi V\over V},
 \label{W1}
\end{equation}
where $\Gamma_{\phi}=\Gamma_{\phi}(\phi)$
is the so-called dissipation coefficient
[cf.~Eq.~(\ref{B3})], responsible for a continuous energy transfer from the
inflaton field $\phi$ to a
radiation bath with energy density $\rho_{\rm R}$.

Assuming quasi-exponential inflation, it follows
\cite{Visinelli}
\begin{equation}
 {\dot{H} \over H^2} \simeq - {\epsilon_\phi \over 1+Q},
\label{W11a}
\end{equation}
\begin{equation}
{\ddot{\phi} \over H\dot{\phi}} \simeq 
	- {1 \over 1+Q} \left( \eta_\phi - \beta_\phi 
	+ {\beta_\phi - \epsilon_\phi \over 1+Q}\right),
\label{W11b}
\end{equation}
\begin{equation}
{\dot{\rho}_{\rm R} \over H\rho_{\rm R}} \simeq 
	- {1 \over 1+Q} \left( 2\eta_\phi - \beta_\phi -\epsilon_\phi
	+ 2{\beta_\phi - \epsilon_\phi \over 1+Q}\right),
\label{W11c}
\end{equation}
where the dissipation ratio $Q$ is defined as
\begin{equation}
 Q \equiv {\Gamma_\phi \over 3H}.
\label{WQ}
\end{equation}

Taking into account that 
$|\dot{H}|/H^2\ll1$ and that the slow-roll approximation requires
$|\ddot{\phi}| \ll |H\dot{\phi}|$
and $|\dot{\rho}_{{\rm R}}| \ll |H\rho_{{\rm R}}|$, from the above
equations we conclude that
\begin{equation}
 \epsilon_\phi \ll 1+Q, \quad 
 |\eta_\phi| \ll 1+Q,   \quad
 |\beta_\phi| \ll 1+Q.
 \label{W12}
\end{equation}
This contrasts with the case of cold inflation, for which $\epsilon_\phi\ll1$
and $|\eta_\phi|\ll1$.

The constants
$c_2$ and $c_3$, arising within the de Sitter swampland
conjectures, Eqs.~(\ref{eq:SC2}) and (\ref{eq:SC3}),
are related to the slow-roll parameters $\epsilon_\phi$ and 
$\eta_\phi$ as $c_2^2<2\epsilon_\phi$ and $c_3<|\eta_\phi|$, respectively,
implying that 
\begin{equation}
 c_2^2 \ll 1+Q, \qquad 
c_3  \ll 1+Q.
\end{equation}

Thus, in the strong dissipative regime of warm inflation, for which $Q\gg1$, 
both $c_2$ and $c_3$ can be of order unity, even if the expansion 
is quasi-exponential, $|\dot{H}|/H^2\ll1$.
This behavior contrasts with the situation in cold inflation, 
for which both $c_2$ and $c_3$ are much smaller than unity during
a quasi-exponential inflationary period.

We now turn to the case of multi-field warm inflation.

Let us start with two scalar fields $\phi$ and $\chi$ with potential
$U(\phi,\chi)$ and the evolution Eqs.~(\ref{B3})--(\ref{B2}).
From the condition that inflation is almost exponential
[see Eq.~(\ref{B5})] and
using Eqs.~(\ref{B1}) and (\ref{B2}), follows that
$\dot{\phi}^2+\dot{\chi}^2+\rho_R \ll U$,
which implies that Eq.~(\ref{B1}) can be written as 
\begin{equation}
 U\simeq 3M_{\rm P}^2 H^2. 
\label{W19}
\end{equation}

Taking the time derivative of the latter expression,
and using Eq.~(\ref{B2}),
we obtain
\begin{equation}
 \dot{\phi} \partial_\phi U + \dot{\chi} \partial_\chi U
 \simeq - 3H \left( \dot{\phi}^2+\dot{\chi}^2+\frac43\rho_R\right), 
\label{W20}
\end{equation}
which can be written as
\begin{equation}
 3H\dot{\phi} \left( 1+ {4\rho_R \over 3\dot{\phi}^2} \right)
 + \partial_\phi U \left( 1 
 	+ {\dot{\chi} \over \dot{\phi}} {\partial_\chi U \over \partial_\phi U}
	+  {3H\dot{\ph} \over \partial_\phi U} {\dot{\chi}^2 \over \dot{\phi}^2}
 \right)
 \simeq0. 
\label{W21}
\end{equation}
Comparing with Eq.~(\ref{B3}),
this is equivalent to neglecting $\ddot{\phi}$ and assuming
\begin{equation}
{\dot{\chi} \over \dot{\phi}} {\partial_\chi U \over \partial_\phi U}
+ {3H\dot{\phi} \over \partial_\phi U} {\dot{\chi}^2 \over \dot{\phi}^2} \simeq0
\label{W22}
\end{equation}
and
\begin{equation}
 {4\rho_R \over 3\dot{\phi}^2} \simeq {\Gamma_\phi \over 3H} ,
 \label{W23}
\end{equation}
yielding
\begin{equation}
 \partial_\phi U \simeq - 3H\dot{\phi} \left( 1+ Q \right).
\label{W24}
\end{equation}

Note that Eq.~(\ref{W22}) can be written as
\begin{equation}
 \partial_\chi U \simeq - 3H\dot{\chi}\, .
 \label{W25}
\end{equation}
which is equivalent to neglecting $\ddot{\chi}$ and 
$\Gamma_\chi \dot{\chi}$ in Eq.~(\ref{B4}),
while Eq.~(\ref{W23}) can be written as
\begin{equation}
 4H\rho_R\simeq\Gamma_\phi \dot{\phi}^2,
 \label{W26}
\end{equation}
which is equivalent to neglecting $\dot{\rho_R}$ and 
$\Gamma_\chi \dot{\chi}^2$ in Eq.~(\ref{Br}).

Now, taking the time
derivative of Eqs.~(\ref{W24})--(\ref{W26}),
we obtain
\begin{equation}
 \dot{\phi} \partial_\phi\Gamma_\phi \simeq \left( 3\dot{H}
 -6H {\ddot{\phi} \over \dot{\phi}}
 +3H {\dot{\rho}_{\rm R} \over \rho_{\rm R}} \right)Q, \label{W27a}
\end{equation}
\begin{equation}
 \partial_{\phi\phi}^2 U 
 + {\dot{\chi} \over \dot{\phi}} \partial_{\phi\chi}^2 U
 \simeq
 -3\dot{H}(1+Q)
 -3H {\ddot{\phi} \over \dot{\phi}}(1-Q)
 -3H {\dot{\rho}_{\rm R} \over \rho_{\rm R}}Q, \label{W27b}
\end{equation}
\begin{equation}
 \partial_{\chi\chi}^2 U 
 + {\dot{\phi} \over \dot{\chi}} \partial_{\phi\chi}^2 U
 \simeq
 -3\dot{H} -3H {\ddot{\chi} \over \dot{\chi}} \,. \label{W27c}
\end{equation}

Following Ref.~\cite{lyth-liddle}, we define the slow-roll parameters
$\epsilon$ and $\eta_{ij}$ as
\begin{equation}
 \epsilon = {1\over2} M_{\rm P}^2\left( 
 {(\partial_\phi U)^2+(\partial_\chi U)^2 \over U^2}\right),
\end{equation}
\begin{equation}
 \eta_{\phi\phi} = M_{\rm P}^2  {\partial_{\phi\phi}^2 U\over U},
\qquad
 \eta_{\phi\chi} = \eta_{\chi\phi} = 
 M_{\rm P}^2  {\partial_{\phi\chi}^2 U \over U},
\qquad
 \eta_{\chi\chi} = M_{\rm P}^2  {\partial_{\chi\chi}^2 U \over U},
\end{equation}
while for $\beta$ we adopt the definition [see Eq.~(\ref{W1})]
\begin{equation}
 \beta = M_{\rm P}^2 {\partial_\phi\Gamma_\phi \over \Gamma_\phi}
 {\partial_\phi U \over U}.
\end{equation}

Using Eqs.~(\ref{W19}), (\ref{W24})--(\ref{W27c}),
it is straightforward to obtain
\begin{equation}
 {\dot{H} \over H^2} \simeq -{\epsilon \over F},
\end{equation}
\begin{equation}
 {\ddot{\phi} \over H\dot{\phi}} \simeq -{1 \over 1+Q}
 \left( \eta_{\phi\phi} + {\dot{\chi} \over \dot{\phi}} \, \eta_{\phi\chi}
 - \beta + {\beta \over 1+Q} - {\epsilon \over F}  \right),
\end{equation}

\begin{equation}
 {\ddot{\chi} \over H\dot{\chi}} \simeq {\epsilon \over F}
 - \eta_{\chi\chi} - {\dot{\phi} \over \dot{\chi}} \eta_{\phi\chi},
\end{equation}

\begin{equation}
 {\dot{\rho}_{\rm R} \over H\rho_{\rm R}} \simeq -{1 \over 1+Q}
 \left( 2\eta_{\phi\phi}
 + 2 {\dot{\chi} \over \dot{\phi}} \, \eta_{\phi\chi}
 - \beta - \epsilon {1+Q \over F}
 + {2\beta \over 1+Q}
 - {2\epsilon \over F} \right),
\end{equation}
where the following notation was introduced,
\begin{equation}
F(Q,\dot{\phi},\dot{\chi})
\equiv
{(1+Q)^2\dot{\phi}^2+\dot{\chi}^2 \over (1+Q)\dot{\phi}^2+\dot{\chi}^2}.
 \label{29}
\end{equation}
Now, taking into account that $|\dot{H}|/H^2\ll1$ and that the 
slow-roll approximation requires 
$|\ddot{\phi}| \ll |H\dot{\phi}|$,
$|\ddot{\chi}| \ll |H\dot{\chi}|$, and 
$|\dot{\rho}_{\rm R}| \ll |H\rho_{\rm R}|$, we obtain
\begin{equation}
 \epsilon \ll F, \quad 
 |\beta|\ll 1+Q, \quad 
 |\eta_{\phi\phi}|\ll 1+Q, \quad
 |\eta_{\chi\chi}|\ll 1, \quad
 \left| {\dot{\chi} \over \dot{\phi}} \eta_{\phi\chi} \right| \ll (1+Q),
 \quad 
 \left| {\dot{\phi} \over \dot{\chi}} \eta_{\phi\chi} \right| \ll 1.
\end{equation}
From these expressions we conclude that,
if $\dot{\chi}^2/\dot{\phi}^2\ll1+Q$, then $F\simeq1+Q$, and,
in the strong dissipative regime of warm inflation, for which $Q\gg1$,
the slow-roll parameters $\epsilon$, $\beta$, and $\eta_{\phi\phi}$
can be of order unity.

On the other hand, the de Sitter swampland conjecture
given by Eq.~(\ref{B14a}), can be written as
\begin{equation}
 c_2^2 \lesssim 2{|\dot{H}| \over H^2} G, \label{W30}
\end{equation}
where 
\begin{equation}
G(Q,\dot{\phi},\dot{\chi}) \equiv 
 {(1+Q)^2\dot{\phi}^2 \over 
 (1+Q)\dot{\phi}^2+\dot{\chi}^2}. \label{W31}
\end{equation}
If $\dot{\chi}^2/\dot{\phi}^2\ll1+Q$, then $G\simeq 1+Q$ and, consequently,
$c_2^2\ll 1+Q$. Therefore, we conclude that, in the strong dissipative 
regime of warm inflation, $c_2$ can be of order unity.
If, however, $\dot{\chi}^2/\dot{\phi}^2 \simeq (1+Q)^2$, then $G\simeq 1$ and,
consequently, $c_2^2\ll 1$. 

The de Sitter swampland conjecture [see Eq.~(\ref{eq:SC3})]
\begin{equation}
M_{\rm P}^2{\partial_{\phi\phi}^{2} U \over U} < -c_3 
\end{equation}
can be written in terms of
the slow-roll parameter $\eta_{\phi\phi}$ as 
$c_3<|\eta_{\phi\phi}|$,	
meaning that $c_3\ll 1+Q$. 	
For $Q\gg1$, corresponding to
the strong dissipative regime of warm inflation, 
$c_3$ can be of order unity.

As in the case of cold inflation, these results can 
be generalized for an arbitrary number of scalar fields,
$\phi$ and $\chi_i$ ($i=1,\dots,N$), with potential
$U(\phi,\chi_1,\dots,\chi_N)$.
As before, the evolution equations yield Eq.~(\ref{CI-f1}),
which, again, \textit{does not} imply 
$\partial_{\chi_i} U +3H\dot{\chi_i}\simeq 0$;
these latter expressions must be assumed,
instead of being derived from Eq.~(\ref{W19}) as in the two-scalar field case.
With this assumption and taking into account that 
$|\dot{H}|/H^2\ll1$ and that the slow-roll approximation requires 
$|\ddot{\phi}| \ll |H\dot{\phi}|$,
$|\ddot{\chi}_i| \ll |H\dot{\chi}_i|$, and 
$|\dot{\rho}_{\rm R}| \ll |H\rho_{\rm R}|$, we obtain
\begin{eqnarray}
 & \epsilon \ll F, \quad 
 |\beta|\ll 1+Q, \quad 
 |\eta_{\phi\phi}|\ll 1+Q, \quad
 \left| {1\over \dot{\chi}_j} \sum_{i=1}^N \eta_{\chi_i\chi_j} \dot{\chi}_i
 \right|\ll 1, \quad \nonumber
\\
 & \left| {1 \over \dot{\phi}} \sum_{i=1}^N \eta_{\phi\chi_i} \dot{\chi}_i
   \right| \ll 1+Q,
 \quad 
 \left| {\dot{\phi} \over \dot{\chi}_i} \eta_{\phi\chi_i} \right| \ll 1,
\end{eqnarray}
where $\eta_{\phi\chi_i}$ and $\eta_{\chi_i\chi_j}$ are slow-roll parameters
defined as
\begin{equation}
 \eta_{\phi\chi_i} = 
 M_{\rm P}^2  {\partial_{\phi\chi_i}^2 U \over U},
\qquad 
 \eta_{\chi_i\chi_j} = M_{\rm P}^2  {\partial_{\chi_i\chi_j}^2 U \over U},
\end{equation}
and $F$ is given by
\begin{equation}
F(Q,\dot{\phi},\dot{\chi_1},\dots,\dot{\chi_N})
 \equiv {(1+Q)^2\dot{\phi}^2+ \sum_{i=1}^N \dot{\chi}_i^2 \over 
(1+Q)\dot{\phi}^2+ \sum_{i=1}^N\dot{\chi}_i^2}.
\end{equation}
Furthermore, $c_2$ is given by Eq.~(\ref{W30}) with
\begin{equation}
G(Q,\dot{\phi},\dot{\chi_1},\dots,\dot{\chi_N})
 \equiv {(1+Q)^2\dot{\phi}^2 \over 
 (1+Q)\dot{\phi}^2+\sum_{i=1}^N\dot{\chi}_i^2}, \label{W31}
\end{equation}
while $c_3$ relates to the slow-roll parameter $\eta_{\phi\phi}$ as 
$c_3<|\eta_{\phi\phi}|$. 
If $\sum_{i=1}^N\dot{\chi}_i^2/\dot{\phi}^2\ll1+Q$, 
then $F\simeq1+Q$ and $G\simeq1+Q$, implying that,
in the strong dissipative regime of warm inflation ($Q\gg1$),
both $c_2^2$ and $c_3$	
can be of order unity.

\section{Discussion and Conclusions\label{sec:concl}}

\phantom{.}\indent In this work we have examined, on quite general grounds,
the possibility of matching the slow-roll conditions of inflation and the
de Sitter
swampland conjectures in the context of cold and warm inflationary models
driven by more than one scalar field.

We have shown that irrespective of the 
number of scalar fields with canonical kinetic energy terms, quasi-exponential
cold inflation and the swampland conjectures are incompatible up to issues
related to the origin of adiabatic curvature perturbations.

The situation is different once dissipation is introduced.
Indeed, in the context of single-field warm inflation, 
it is known that the slow-roll conditions and the swampland conjectures 
can be matched provided the dissipation ratio,
$Q=\Gamma_{\ph}/(3H)$, satisfies the condition $Q\gg1$,
corresponding to a strong dissipative regime. 
Even though the situation of warm inflation driven by two scalar fields is
more complex as it requires more parameters in order to fully characterize
the slow-roll conditions, we have shown that quasi-exponential inflation and
de Sitter swampland conditions can be reconciled provided 
a strong dissipative regime is ensured for one of the scalar fields,
say $\phi$, and the other scalar field $\chi$ satisfies
the condition $\dot{\chi}^2/\dot{\phi}^2\ll1$.
For more than two scalar fields, as expected, the slow-roll conditions are
richer, however, they can be made compatible with the swampland conditions,
likewise in the cases of single- and two-field
inflation, if the strong dissipative regime 
holds for one of the fields, say $\phi$, and if
the condition $\sum_{i=1}^N\dot{\chi}_i^2/\dot{\phi}^2\ll1+Q$
is satisfied by the remaining fields,
$\chi_i$. However, differently from the two-scalar-field case,
the conditions $\partial_{\chi_i} U +3H\dot{\chi}_i\simeq 0$ should be 
imposed, since they do not follow from the slow-roll expression
$U\simeq 3M_{\rm P}^2 H^2$.

To close the discussion we emphasize that our work opens the possibility of
reconciling the swampland conjectures with warm inflation, provided the 
single- or multi-field models admit, during the slow-roll, a strong dissipative
regime and satisfy
the conditions discussed above.
These extra requirements might be relevant to narrow down
the possible class of viable effective models that, although not in the
landscape of string theory, are suitable from the point of view of inflation.
It is intriguing that inflation with dissipative features is also an important
ingredient in a recent proposal to address the cosmological constant
problem \cite{Bertolami2021}. 
Moreover, it is quite interesting that dissipation, being a quite generic
manifestation of the arrow of time at the macroscopic level, appears in a
fundamental discussion about the viable emerging solutions of string theory.  

\vspace{0.5cm}

{\bf Acknowledgments}

\noindent
PMS acknowledges support from Funda\c{c}\~ao para a Ci\^encia e a
Tecnologia (Portugal) through the research grants UIDB/04434/2020
and UIDP/04434/2020.

\end{document}